\documentclass[11pt]{article}
\usepackage{moriond,epsfig}

\def\Journal#1#2#3#4{{#1} {\bf #2}, #3 (#4)}

\def\NPB{{\em Nucl. Phys.} B}
\def\PLB{{\em Phys. Lett.}  B}
\def\PRL{\em Phys. Rev. Lett.}
\def\PRD{{\em Phys. Rev.} D}

\def\be{\begin{equation}}
\def\ee{\end{equation}}
\def\bea{\begin{eqnarray}}
\def\eea{\end{eqnarray}}

\def\la{\mathrel{\mathpalette\fun <}}
\def\ga{\mathrel{\mathpalette\fun >}}
\def\fun#1#2{\lower3.6pt\vbox{\baselineskip0pt\lineskip.9pt
  \ialign{$\mathsurround=0pt#1\hfil##\hfil$\crcr#2\crcr\sim\crcr}}}
\begin{document}
\vspace*{4cm}
\title{Gravitational Effects on Particle Dark Matter and Indirect Detection}

\author{G\"unter Sigl, Gianfranco Bertone}

\address{GReCO, Institut d'Astrophysique de Paris, C.N.R.S.,
98 bis boulevard Arago, F-75014 Paris, France}

\maketitle\abstracts{
The annihilation signal of particle dark matter can be strongly
enhanced in over-dense regions such as close to the Galactic centre.
We summarize some of our recent results on fluxes of $\gamma-$rays,
neutrinos and radio waves under different assumptions for the
largely uncertain dark matter profile which close to the Galactic
centre is strongly influenced by the super-massive black hole. We
apply this to two particle dark matter scenarios, namely the
case of neutralinos in supersymmetric scenarios and the lowest
Kaluza-Klein excitation of the hyper-charge gauge boson in scenarios
where the Standard Model fields propagate in one microscopic
extra dimension.}

\section{Introduction}
Over the last thirty years, independent pieces of evidence have 
accumulated in favor of the existence of \textit{Dark Matter} (DM), 
indicating that most of the matter of the universe is non baryonic and of
unknown nature. Particle physicists have come up with various 
DM candidates. The most promising and most extensively studied is the 
so-called \textit{neutralino}, the Lightest Supersymmetric Particle (LSP)
which arises in supersymmetric models and is stable in models with
conserved R-parity~\cite{jkg}.

Although theoretically well motivated, 
\textit{neutralinos} are not the only viable DM candidates.
For instance, models with compact extra dimensions possess
plenty of new states, Kaluza--Klein (KK) particles.
In models with \textit{Universal Extra Dimensions}~\cite{Appelquist:2000nn},
in which all Standard Model fields propagate in extra dimensions, including 
fermions, the Lightest Kaluza-Klein Particle (LKP) can be stable and
was recently shown to be a viable DM 
candidate~\cite{Servant:2002aq}. Precision electroweak measurements
put a lower limit of $M\ga 300\,$GeV on the mass of the
LKP~\cite{Appelquist:2000nn} in the case of one universal extra
dimension.

Indirect signatures of DM are typically proportional to the line
of sight integral of the squared density. Unfortunately, there is
still no consensus about the shape of 
dark matter halos. High-resolution N-body simulations suggest the
existence of ``cusps'', with the inner part of the halo density
following  a power law $\propto r^{-\gamma}$ in the distance $r$ to
the Galactic centre (GC) with index $\gamma$ possibly as high as 1.5,
but it could also be $\simeq1.0$ or shallower~\cite{swsty}.
On the other hand observations of rotation curves of galaxies
seem to suggest much shallower inner profiles~\cite{salucci},
but other groups claim the impossibility of constraining dark 
matter with such observations~\cite{vandenBosch:2000rz}).

The usual parametrisation for the dark matter halo density is
\begin{equation}
  \rho(r)= \frac{\rho_0}{(r/R)^{\gamma}
  [1+(r/R)^{\alpha}]^{(\beta-\gamma)/\alpha}}\label{profile}
\end{equation}
In Tab.~\ref{tab} we give the values of the respective parameters for
some of the most widely used profile models.

\begin{table}
\caption[...]{\label{tab} Parameters of some widely used profile models,
namely the Kravtsov et al. (Kra, \cite{kra}), 
Navarro, Frenk and White (NFW, \cite{Navarro:1995iw}), Moore et al. 
(Moore, \cite{Moore:1999gc}) and modified isothermal (Iso, e.g. 
\cite{Bergstrom:1997fj}) profiles for 
the dark matter density in galaxies in Eq.~(\ref{profile}).}
\begin{center}
\begin{tabular}{cccccc}
&$\alpha$&$\beta$&$\gamma$&R (kpc)\\ \hline \\
Kra& 2.0& 3.0&0.4 & 10.0 \\
NFW& 1.0& 3.0& 1.0& 20 \\
Moore& 1.5& 3.0& 1.5& 28.0 \\
Iso& 2.0& 2.0& 0& 3.5 \\
\end{tabular}
\end{center}
\end{table}

Observations of the velocity dispersion of high proper motion stars suggest
the existence of a super-massive black hole lying at the centre of our
Galaxy, with a mass $\approx 2.6 \times 10^6 M_\odot$ ~\cite{Ghez:1998ab}. 
It has been argued ~\cite{Gondolo:1999ef} that the process of adiabatic 
accretion of dark matter on this central super-massive black hole would
in addition produce a ``spike''
in the dark matter density profile, leading to a power law index possibly as
high as $\gamma \approx 2.4$. Although central spikes could be destroyed
by astrophysical processes such as hierarchical mergers 
~\cite{Ullio:2001fb,Merritt:2002vj}, these dynamical destruction 
processes are unlikely to have occurred for the Milky
Way~\cite{Bertone:2002je}. The existence of such spikes would produce
a dramatic enhancement of the annihilation radiation from the GC, and
would allow to put stringent constraints on dark matter particles
properties and
distributions~\cite{Gondolo:1999ef,Gondolo:2000pn,Bertone:2001jv,Bertone:2002je}.
For an initially power-law type profile of index
$\gamma$, as predicted by high resolution N-body
simulations~\cite{Navarro:1995iw},
the corresponding dark matter profile after accretion is steepened
to an index $\gamma_{sp}=(9-2\gamma)/(4-\gamma)$ within ~$10^{-5}\,$pc
of the central black hole.
For more details see~\cite{Gondolo:1999ef}).

The observable annihilation flux of a secondary particle of type $i$
(we focus on  $\gamma-$rays and neutrinos) can be written as
\begin{equation}
\Phi_i(\psi,E)=\sigma v \frac{dN_i}{dE} \frac{1}{4 \pi M^2}
\int_{\mbox{line of sight}}d\,s
\rho^2\left(r(s,\psi)\right)\,,\label{flux}
\end{equation}
where the coordinate $s$ runs along the line of sight, in a
direction making an angle $\psi$ respect to the direction
of the GC. $\sigma v$ is the annihilation cross 
section, $dN_i/dE$ is the spectrum of secondary particles
per annihilation, and $M$ is the DM particle mass.

The particle physics parameters $\sigma v$ and $dN_i/dE$ depend
on the model. For neutralino DM we use the DarkSUSY~\cite{darksusy}
tool to select parameter values resulting in acceptable DM
relic densities and being consistent with accelerator constraints.
The LKP scenario has only one independent parameter, namely the
LKP mass or, equivalently, the linear size of the extra dimension.
The cross sections and branching ratios are then taken
from~\cite{Servant:2002aq}. The calculation of $dN_i/dE$ in general
involves three steps: Determination of the partial cross sections
into lepton and quark final states, fragmentation of the quark
final states into hadrons and decay of the hadrons into stable
end products, namely nucleons, $e^\pm$, $\gamma-$rays, and neutrinos.
For more details on the calculation
of the secondary spectra $dN_i/dE$ see~\cite{bss}.

\section{Gamma-Ray and Neutrino Fluxes}
\begin{figure}[ht]
\psfig{figure=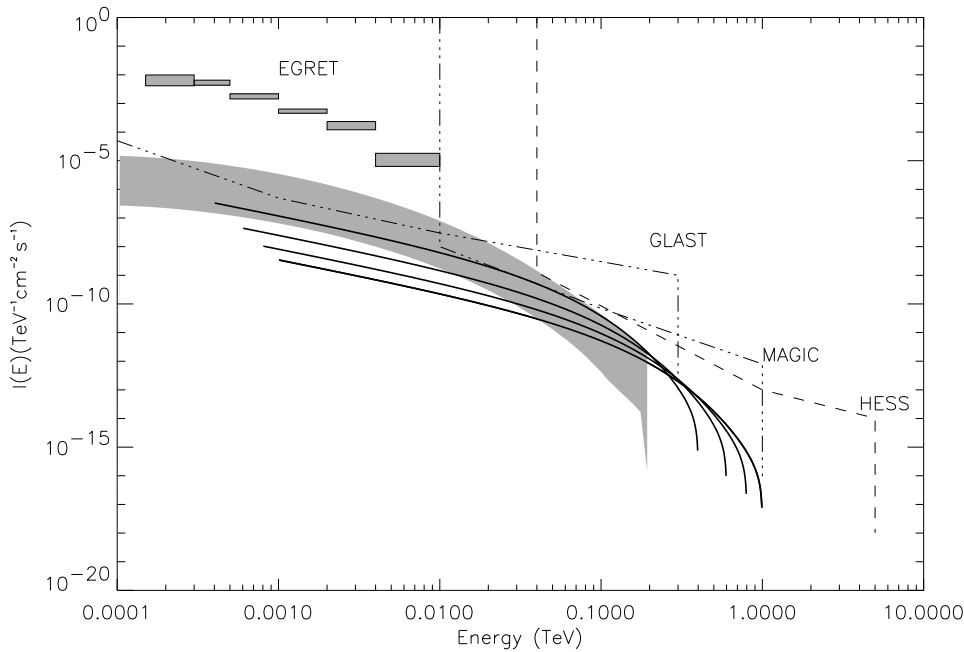,width=0.8\textwidth}
\caption[...]{Expected $\gamma-$ray fluxes from the GC for the LKP scenario
for (lines top to bottom) $M=0.4$, 0.6, 0.8,
and 1 TeV, assuming an NFW profile without accretion spike. The shaded
area shows typical $\gamma-$ray fluxes predicted for
neutralinos of mass $\simeq200\,$GeV. Also shown are EGRET data and expected 
sensitivities of the future GLAST, MAGIC and HESS experiments.}
\label{fig1}
\end{figure}

In Fig.~\ref{fig1} we show the $\gamma-$ray flux
in a solid angle $\Delta\Omega=10^{-3}$ in the direction
of the GC predicted by Eq.~(\ref{flux}) for some of the DM models
discussed above, assuming
a NFW profile without accretion spike. In the same figure we show
for comparison observational data from 
EGRET \cite{mayer}, and expected sensitivities of the future experiments 
GLAST \cite{Sadrozinski:wu}, MAGIC \cite{Petry:1999fm} and 
HESS~\cite{Volk:2002iz}.

\begin{figure}[ht]
\psfig{file=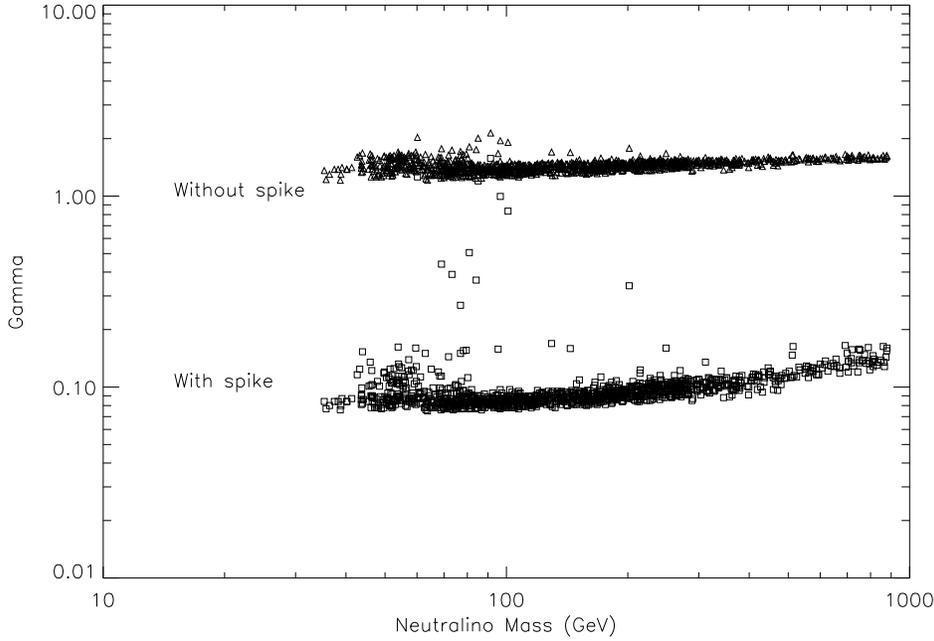,width=0.8\textwidth}
\caption{Required value of $\gamma$ to reproduce the EGRET GC $\gamma-$ray
flux for a typical set of supersymmetric DM models. Squares
and triangles are for a halo model respectively with or without a
central accretion spike.}
\label{fig2}
\end{figure}

As can be seen from Fig.~\ref{fig2}, to explain the EGRET GC $\gamma-$ray
flux would require DM slopes of $\gamma\simeq1.5$ if no accretion
spike is present. Presence of an accretion spike would instead require
DM slopes $\gamma\simeq0.1$ further out.

The neutrino fluxes are in general comparable to the $\gamma-$ray
fluxes, but for for a NFW profile without spike they are typically
a factor $\sim10^3$ below the sensitivity of experiments such as
ANTARES and the Sun turns out to be a more easily detectable neutrino
source in this case.

\section{Synchrotron Radiation}

Another interesting mean of indirect DM detection is the synchrotron
radiation originated from the propagation of secondary $e^{\pm}$ in
the Galactic magnetic field. 

The magnetic field is supposed to be at equipartition (for details 
see \cite{melia,Bertone:2001jv}) in the inner part of the 
Galaxy and constant elsewhere. More specifically
\begin{equation}
B(r) = \mbox{max}\left[ 324 \mu \mbox{G} \left(\frac{r}{\mbox{pc}}\right)^
{-5/4}, 6\mu\mbox{G}\right]\,,\label{mag}
\end{equation}
which means that the magnetic field is assumed to be in equipartition
with the plasma out to a galactocentric distance $r_c=0.23\,$pc,
and to be equal to a typical value observed throughout the Galaxy
at larger distances.

Smaller magnetic fields away from the central region would imply a shift of
the radio spectrum to lower energies and thus, in the range of frequencies 
we are interested in, a higher flux for a given frequency. 
This would also translate into stronger constraints on DM mass and 
annihilation cross section, thus Eq.~(\ref{mag}) is likely to be
conservative. Note that magnetic fields
stronger than equipartition values are physically unlikely.

The energy loss of $e^\pm$ is dominated by synchrotron radiation
for which the loss time is much shorter than the diffusion length
in magnetic fields of the order of Eq.~(\ref{mag})~\cite{Bertone:2001jv}.
The synchrotron flux per solid angle at a given frequency $\nu$
(cf. Eq.~(22) in~\cite{Bertone:2001jv}) can then be approximated by
\begin{equation}
L_{\nu}(\psi) \simeq \frac{1}{4 \pi}\frac{9}{8}
\left(\frac{1}{0.29 \pi} \frac{ m_e^3 c^5}{e} 
\right)^{1/2} \frac{\sigma v}{M^2} Y_e(M,\nu)  \;\nu^{-1/2}\;
\int_0^{\infty} ds \;\; \rho^2\left(r(s,\psi)\right) B^{-1/2}
\left(r(s,\psi)\right)\,,\label{synchrolum}
\end{equation} 
where $s$ is the coordinate running along the line of sight. $Y_e(M,\nu)$ is
the average number of secondary electrons above the energy $E_m(\nu)$
at which synchrotron emission peaks at frequency $\nu$ for magnetic
field strength $B$. For $r < r_c$,
\begin{equation}
E_m(\nu)=\left( \frac{4 \pi}{3} \frac{m_e^3 c^5}{e} \frac{\nu}{B}
\right)^{1/2}\simeq 0.3 \left(\frac{\nu}{400\,{\rm MHz}}\right)^{1/2}
\left(\frac{r}{\mbox{pc}}\right)^{5/8} \mbox{GeV}\,,\label{ecritic}
\end{equation}
which at the inner edge of the profile, corresponding to the Schwarzschild 
radius of the GC black hole, $r_s=1.3\times10^{-6}$ pc, takes the value
$E_m(400\,\mbox{MHz})\simeq2.2 \times 10^{-5}$ GeV. Since, therefore,
$E_m(400\,\mbox{MHz}) \ll M$ always, basically all secondary
$e^\pm$ from DM annihilation are produced above this energy and contribute
to the radio flux. Their number $Y_e(M)\simeq5$ is again calculated from
the specific particle physics model.

\begin{figure}[ht]
\psfig{file=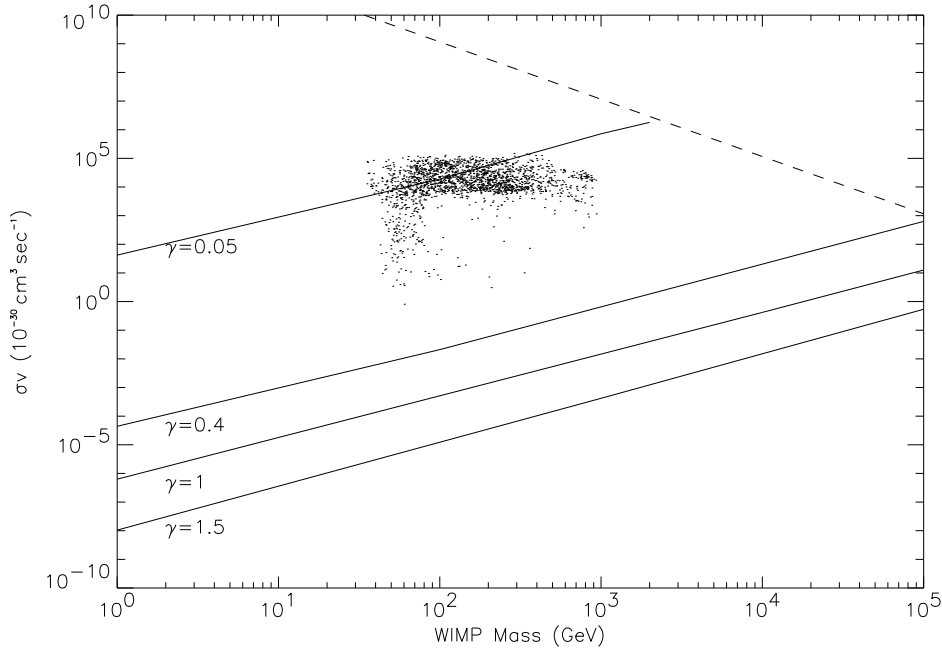,width=0.8\textwidth}
\caption[...]{Exclusion plot based on the comparison between predicted and
observed radio flux at 408 MHz from the inner 4 arcmin around the GC,
assuming the
presence of an accretion spike. The solid lines mark the upper limits
on the annihilation cross section for various DM profiles $\gamma$
outside the accretion spike. Dots represents a scan of SUSY neutralino
DM and the dashed line represents the unitarity bound for s-channel
annihilation, $\sigma v\la M^{-2}$.}
\label{fig3}
\end{figure}

If an accretion spike is present, Eq.~(\ref{synchrolum}) is significantly
modified by synchrotron self absorption whose effect we have estimated
in~\cite{Bertone:2001jv}. In the absence of a spike self absorption is
negligible. One can now compare the predicted radio flux
with the one observed at 408 MHz in a cone of half-width
4 arcsec centered on the GC, which is $\la 0.05\,$Jy~\cite{davies}.
This yields the constraints on the annihilation
cross section shown in Fig.~\ref{fig3} if an accretion spike is
present. In this case most neutralino models would be ruled out
but for the most shallow DM profiles~\cite{Gondolo:2000pn} $\gamma\la0.05$.

In contrast, for a NFW profile without accretion spike one obtains
the constraint
\begin{equation}
\sigma v\la1.5\times 10^{-26}\left(\frac{M}{100\mbox{\small{GeV}}} \right)^2 
\frac{5}{Y_e(M)}\, \mbox{cm}^{3} \mbox{s}^{-1}\,.
\end{equation}
This is at the high end of typical neutralino cross sections, see
Fig.~\ref{fig3}.

\begin{figure}
\psfig{file=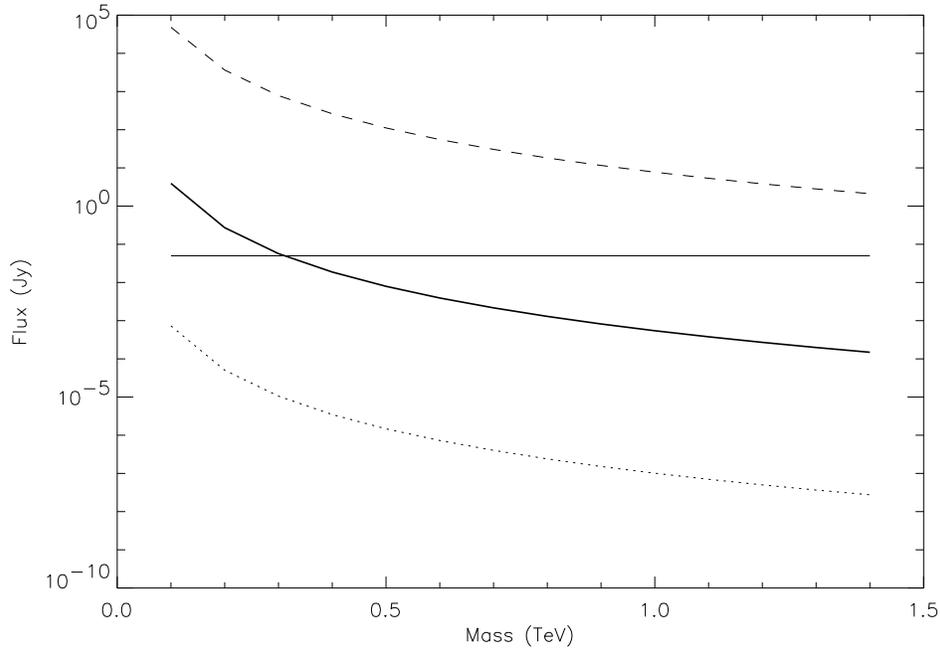,width=0.8\textwidth}
\caption{Radio flux from a 4 arcsec cone around the GC predicted
in the LKP scenario as a function of the LKP mass for different
density profiles. Bottom to top: Kravtsov et al.
(dotted line), NFW (solid line) and Moore (dashed line) profiles.}
\label{fig4}
\end{figure}

Constraints on the LKP scenario for various profiles without accretion
spike are shown in Fig.~\ref{fig4}. For a NFW profile, for example,
the lower limit $M\ga300\,$GeV results.

We also compared predicted and observed radio fluxes at high latitude.
The strongest constraints result from
the lowest frequencies at which free-free and synchrotron self-absorption
are not yet important, i.e. around $10\,$MHz~\cite{cane}. Here,
the observed background
emission between $0^\circ$ and $90^\circ$ from the Galactic
anti-centre is $\simeq6\times10^6\,$Jy. Comparing with the
predicted emission results in the limit
\begin{equation}
\sigma v \la10^{-24}\left(\frac{M}{100\mbox{\small{GeV}}} \right)^2 
\frac{Y_e(1\mbox{\small{TeV})}}{Y_e(M)}\,\mbox{cm}^{3} \mbox{s}^{-1}. 
\end{equation}
While this is considerably weaker than the constraints above,
it is largely independent of the unknown GC dark
matter profile.

\section{Conclusions}
We have discussed annihilation signals of dark matter from the Galactic centre
and halo in the context of supersymmetric neutralino dark matter
and Kaluza Klein states as dark matter in scenarios where Standard
Model gauge boson fields propagate in one extra dimension. The largest
uncertainties come from the unknown dark matter profile close to
the Galactic centre.
Comparison of Fig.~\ref{fig2} and \ref{fig4} shows that currently,
assuming Eq.~(\ref{mag}) for the magnetic field close to the Galactic
centre, the synchrotron flux gives stronger constraints than the
$\gamma-$ray flux. When instruments such as GLAST, MAGIC, and HESS
will turn on, the $\gamma-$ray channel will provide constraining power
comparable to the synchrotron channel, with the advantage that
the $\gamma-$ray fluxes do not depend on the magnetic field uncertainties,
although they may partly be absorbed. Our results on $\gamma-$rays are
consistent
with the findings in~\cite{swsty} for a NFW profile without accretion
spike which leads to fluxes marginally detectable with next generation
$\gamma-$ray detectors. There it has also been shown that the clumpy
structure of the dark matter in our Galaxy does not significantly
strengthen the constraints. Finally, constraints based on the
synchrotron emission may be enhanced by more systematically searching
for directions in the sky where the ratio of predicted to observed
radio flux is maximized.

\section*{Acknowledgments}
This is based on work in collaboration with Joseph Silk and
Geraldine Servant.

\end{document}